\documentclass[12pt]{article}
\usepackage{amssymb,amsmath,epsfig}
\usepackage{graphicx}
\allowdisplaybreaks

\begin{document}
\title{\bf Analysis of Cosmic Evolution admitting Garcia-Salcedo Ghost
and Generalized Ghost Dark Energy Models}
\author{M. Zeeshan Gul \thanks{mzeeshangul.math@gmail.com}~,
M. Sharif \thanks {msharif.math@pu.edu.pk}~ and~ I. Hashim
\thanks{imran.hashim@math.uol.edu.pk}\\
Department of Mathematics and Statistics, The University of
Lahore,\\
1-KM Defence Road Lahore-54000, Pakistan.}
\date{}
\maketitle

\begin{abstract}
This study aims to explore the Garcia-Salcedo ghost dark energy and
generalized ghost dark energy models in the context of
$f(\mathrm{R},\mathrm{T^2})$ theory, where $\mathrm{R}$ is the Ricci
scalar and $\mathrm{T^2}$ is the self-contraction of stress-energy
tensor. We investigate the non-interacting case only corresponding
to flat Friedmann-Robertson-Walker universe. We reconstruct the
corresponding $f(\mathrm{R},\mathrm{T^2})$ gravity models using the
considered dark energy models by taking two particular models of
this gravity. The stability analysis is performed for all the cases.
The behavior of the equation of state parameter is also checked. It
is found that some of the reconstructed $f(\mathrm{R},\mathrm{T^2})$
models successfully describe both the phantom and quintessence
epochs of the universe, which support the current cosmic accelerated
expansion. This exploration reveals the intricate connections
between dark energy models and modified gravitational theory,
providing significant understanding into the dynamics of the cosmos
on a large scale.
\end{abstract}

\section{Introduction}

Various cosmic observations such as supernova type-Ia, cosmic
microwave background radiations and large-scale structures indicate
that the universe is currently expanding. Scientists believe that
this expansion is caused by a mysterious force called dark energy
(DE), which is characterized by a large negative pressure. Several
alternative techniques have been formulated to comprehend the impact
of DE on the ongoing cosmic acceleration. Recently, researchers
introduced a new DE model called ghost dark energy (GDE) model,
based on the dynamics of quantum chromodynamics (QCD) Veneziano
ghost fields \cite{2}-\cite{2d}. This model has notable physical
characteristics that contribute to the expanding universe and the
Veneziano ghost field resolves the $U(1)$ problem. In the context of
curved space, the GDE model contributes to the vacuum energy density
as $\rho_{DE}\approx \Lambda_{QCD}^{3} H \approx (10^{-3}eV)^{4}$.
This density plays a crucial role in determining the current cosmic
expansion, where $H$ represents the Hubble parameter, $\rho_{DE}$
denotes the density of DE, $\Lambda_{QCD}\approx100 MeV$ and
$H\approx10^{-33}eV$. According to Zhitnitsky, the vacuum energy
related to the ghost field can be represented as
$H+\mathcal{O}(H^2)$ \cite{3}. The additional term plays a crucial
role during the early stages of the cosmic evolution and it could be
a manifestation of DE. He argued that taking into account the $H^2$
term in the energy density could lead to better agreement with
observational data as compared to GDE.

Einstein's theory of gravity revolutionized our understanding of
gravity offering profound insights into cosmic phenomena and aligns
with the solar system tests. However, significant challenges persist
with general relativity (GR) including the presence of singularities
in the early universe and inside the black holes \cite{4,4.a}. To
address such issues, modifications to GR have been proposed. One of
the simplest extensions is known as the $f(\mathrm{R})$ theory,
which introduces a function of the Ricci scalar $(\mathrm{R})$ in
the Einstein-Hilbert action \cite{5}. A comprehensive understanding
of this theory and its implications can be found in
\cite{6}-\cite{6.2}. Harko et al \cite{6b.1} generalized the
modified $f(\mathrm{R})$ theory by including the trace of the
energy-momentum tensor in the functional action, named as
$f(\mathrm{R},\mathrm{T})$, which provides novel insights into
gravitational phenomena. Haghani et al \cite{6b.8} further extended
the $f(\mathrm{R},\mathrm{T})$ theory by introducing the additional
term $Q=\mathrm{R}_{\mu\nu}\mathrm{T}^{\mu\nu}$ in the geometric
part of the action, known as $f(\mathrm{R},\mathrm{T},\mathrm{Q})$
theory. There is a growing interest among researchers in
investigating this theory for multiple reasons including its
theoretical consequences, alignment with observational data and its
importance in cosmic scenarios \cite{6b.9}-\cite{6b.14}. The
significant literature and implications of
$f(\mathrm{R},\mathrm{T})$ modified theory have been examined in
\cite{6b.2}-\cite{6b.4}. Scientists have explored various extended
gravitational theories beyond $f(\mathrm{R})$ to unravel the
mysteries of the cosmos \cite{6a.1}-\cite{6a.5}.

One such modification is known as $f(\mathrm{R},\mathrm{T^2})$ or
energy-momentum squared gravity (EMSG), which introduces a nonlinear
term $(\mathrm{T^2}=\mathrm{T}^{\mu\nu}\mathrm{T}_{\mu\nu})$ in the
functional action \cite{7}. This modified proposal reduces to GR in
vacuum and its deviation becomes apparent in the high-curvature
regime. Incorporating the terms involving self-contraction of the
energy-momentum tensor, this theory explores intriguing cosmological
implications diverging from GR in the presence of a matter source.
This modified approach offers a versatile framework for probing
fundamental aspects of gravity and cosmology. The energy-momentum
squared gravity exhibits a maximum energy density with a minimal
scale factor in the early universe, which ensures the presence of a
nonsingular bounce in this theory \cite{8}. Board and Barrow
\cite{11} studied cosmic evolution through exact cosmological
solutions in EMSG theory. Moraes and Sahoo \cite{9} conducted a
study on wormhole solutions using non-exotic matter in the same
theory to understand theoretical possibilities of traversable
shortcuts in the universe. Additionally, the same theoretical
framework was used to investigate physical characteristics of
compact stars \cite{10}. Barbar et al \cite{12} discussed the
feasibility and implications of a bouncing cosmological scenario in
this theory. Sharif et al \cite{13}-\cite{17.a} investigated various
astrophysical and cosmological aspects of this theory.

The primary motivation behind investigating the Garcia-Salcedo GDE
and GGDE models in modified $f(\mathrm{R},\mathrm{T}^2)$ theory lies
in understanding the accelerated expansion of the universe, which
has been one of the most profound discoveries in recent decades.
These theoretical DE models are significant to explain the cosmic
acceleration. By incorporating them into the modified
$f(\mathrm{R},\mathrm{T}^2)$ framework, we can explore new aspects
of these models and test their viability against observational data.
This investigation allows us to examine the theoretical consistency
of DE models in a modified context. By embedding the Garcia-Salcedo
GDE and GGDE models in $f(\mathrm{R},\mathrm{T}^2)$ gravity, we can
derive new insights into cosmic implications, leading to prediction
of new cosmological behavior or corrections to existing ones. This
broader understanding of DE dynamics aims to deepen our
understanding of DE, address existing theoretical challenges and
explore new cosmological models that could better align with
observational data.

Reconstruction techniques used in modified gravity theories provide
a systematic approach to formulate viable DE models. By specifying
the evolution of the scale factor, one can obtain the corresponding
gravitational theories or DE properties that are consistent with the
observational data. This method enhances the physical validity and
capability of the resulting models for explaining the accelerated
cosmic expansion. Copeland et al \cite{18} investigated various
methods to shed light on the accelerated expansion of the cosmos.
Amendola et al \cite{19} studied the conditions for cosmological
viability in $f(\mathrm{R})$ gravity regarding DE models. Sheykhi
and Movahed \cite{20} analyzed various cosmological parameters to
comprehend the expansion of the cosmos by examining the  GDE model
corresponding to flat FRW universe. Odintsov et al \cite{21}
investigated specific $f(\mathrm{R},G)$ gravity models ($G$ is the
Gauss-Bonnet invariant) to determine the effective scenarios for DE
and inflationary epochs. Chaudhary et al \cite{22} examined the
dynamics of $f(\mathrm{R},G,\mathrm{T})$ model through DE models.
Akarsu et al \cite{16} explored gravitational wave radiation and
relativistic compact binaries. We have studied the GDE in the
background of $f(\mathrm{R},\mathrm{T^2})$ theory for the
non-interacting case \cite{23}. Our investigations uncovered an
intriguing phenomenon, offering insights into the complex interplay
between DE and gravitational theory.

Cai and Tuo \cite{22a} explored the QCD model in a flat FRW universe
and found that their results were consistent with observational
data. Sheykhi et al \cite{24} studied the implications of the GDE
model in Brans-Dicke theory and discovered that this model crosses
the phantom line with appropriate parameters. Karami et al \cite{25}
analyzed cosmic evolution by examining various cosmological
parameters in the background of the generalized GDE (GGDE) model.
Zubair and Abbas \cite{26} reconstructed the
$f(\mathrm{R},\mathrm{T})$ theory for QCD models and demonstrated
that the reconstructed function exhibits both phantom and
quintessence regimes in a flat FRW universe model. The FRW universe
model offers a coherent framework to comprehend the expansion and
evolution of the cosmos. The cosmological solutions through
gravitational decoupling approach in modified theory has been
discussed in \cite{26.a,26.b}. Recently, the study of relativistic
models in modified gravities has been discussed in
\cite{27}-\cite{27d}. Sharif and Ajmal \cite{27e} reconstructed the
$f(Q)$ theory in the framework of the GGDE model and noted that the
reconstructed GGDE $f(Q)$ model is stable and aligns with the
prevailing understanding of cosmic accelerated expansion.

Our study is focused on reconstructing the
$f(\mathrm{R},\mathrm{T^2})$ models using Gracia-Selcedo GDE  and
GGDE models with flat FRW universe model. We analyze the
evolutionary pattern and stability of the reconstructed models. The
paper is structured as follows. In section \textbf{2}, we present
the modified field equations of $f(\mathrm{R},\mathrm{T^2})$ theory.
Sections \textbf{3} and \textbf{4} describe the reconstruction of
the $f(\mathrm{R},\mathrm{T^2})$ models through Garcia-Salcedo GDE
and GGDE models, respectively, and investigate the evolutionary
behavior. Finally, we summarize our key findings in section
\textbf{5}.

\section{Energy-Momentum Squared Gravity}

The extension of GR in the realm of $f(\mathrm{R},\mathrm{T^2})$
gravity represents a significant advancement in the field of
theoretical physics. It provides a comprehensive framework for
exploring gravitational phenomena beyond standard models. In this
modified theory, the action includes terms involving Ricci scalar
and self-contraction of the energy-momentum tensor. By adding the
contraction term, this theory offers a richer understanding of the
complex relationship between matter and spacetime. Through rigorous
theoretical analysis and observational testing, this modification
holds the promise of deepening our understanding of fundamental
cosmological principles and potentially unveiling new insights into
the essence of the cosmos. The corresponding action is expressed as
\cite{7}
\begin{equation}\label{1}
\mathrm{S}=\int d^{4}x\bigg(\frac{1}{2\kappa^{2}}
f(\mathrm{R},\mathrm{T^2})+\mathbb{L}_{m}\bigg)\sqrt{-g}.
\end{equation}
Here, $\mathbb{L}_{m}$ signifies the matter-Lagrangian density, $g$
represents determinant of the metric tensor and $\kappa^{2}=1$ is
the coupling constant. The corresponding field equations are derived
by varying this action with respect to the metric tensor as
\begin{equation}\label{2}
f_{\mathrm{T^2}}\Theta_{\mu\nu}+\mathrm{R}_{\mu\nu}f_{\mathrm{R}}-
\frac{1}{2}g_{\mu\nu}f+(g_{\mu\nu}\Box-
\nabla_{\mu}\nabla_{\nu})f_{\mathrm{R}}=T_{\mu\nu},
\end{equation}
where
\begin{equation}\label{3}
\Theta_{\mu\nu}=-4\frac{\partial^{2}\mathbb{L}_{m}}{\partial
g^{\mu\nu}\partial g^{\varphi\beta}}T^{\varphi\beta}-2\mathbb{L}_{m}
(T_{\mu\nu}-\frac{1}{2}g_{\mu\nu}\mathrm{T})-\mathrm{T}T_{\mu\nu}+2
T^{\varphi}_{\mu} T_{\nu\varphi}.
\end{equation}
Here, $f_{\mathrm{R}}=\frac{\partial f}{\partial\mathrm{R}}$,
$f_{\mathrm{T^2}}=\frac{\partial f}{\partial\mathrm{T^2}},$
$\Box=\nabla^{\mu}\nabla_{\mu}$ and $\nabla_{\mu}$ is the covariant
derivative.

We consider the isotropic matter distribution as
\begin{equation}\label{4}
T_{\mu\nu}=(\rho+P)\mathbb{U}_{\mu}\mathbb{U}_{\nu}-Pg_{\mu\nu},
\end{equation}
where $\mathbb{U}_{\mu}$ represents the four velocity, $\rho$ is the
energy density and $P$ is the fluid pressure. The matter-Lagrangian
plays a crucial role in different cosmic phenomena as it defines the
matter configuration in spacetime. When dealing with isotropic
matter configuration, the specific expression for the
matter-Lagrangian density can offer valuable insights. In the
literature, one common formulation of matter-Lagrangian is
$\mathbb{L}_m=P$ \cite{29}, which has been considered in various
theoretical frameworks and significantly explains the phenomenon
associated with uniform matter configuration in the context of
cosmology. As a result, this matter-Lagrangian density provides a
physically significant and mathematically accessible framework for
studying uniform matter arrangements in gravitational physics.
Equation (\ref{3}) corresponding to $\mathbb{L}_m=P$ turns out to be
\begin{equation}\label{5}
\Theta_{\mu\nu}=-(\rho^{2}+3P^{2}+4
P\rho)\mathbb{U}_{\mu}\mathbb{U}_{\nu}.
\end{equation}
Rearranging Eq.\eqref{2}, we have
\begin{eqnarray}\label{6}
G_{\mu\nu}=\frac{1}{f_{\mathrm{R}}}(T_{\mu\nu}+T_{\mu\nu}^{GDE}),
\end{eqnarray}
where
\begin{equation}\label{7b}
T^{GDE}_{\mu\nu}=\frac{1}{2}g_{\mu\nu}(f-
\mathrm{R}f_{\mathrm{R}})-(g_{\mu\nu}\Box
-\nabla_{\mu}\nabla_{\nu})f_{\mathrm{R}}-
f_{\mathrm{T^2}\Theta_{\mu\nu}}.
\end{equation}

To examine the mysterious universe, we assume the flat FRW model as
\begin{equation}\label{7c}
ds^{2}=d\mathrm{t}^{2}-{a}^{2}(\mathrm{t})d\chi^{2}.
\end{equation}
Here, $d\chi^{2}=dx^2+dy^2+dz^2$ and ${a}(\mathrm{t})$ determines
the scale factor. Using Eqs.(\ref{2})-(\ref{7c}), the resulting
field equations are
\begin{eqnarray}\label{7d}
3H^{2}&=&\rho_{total},
\\\label{7e}
3H^{2}+2\dot{H}&=&-P_{total},
\end{eqnarray}
where $H=\frac{\dot{a}}{a}$ is the Hubble parameter (dot means
derivative with respect to time), $\rho_{total}=\rho+\rho_{GDE}$,
$P_{total}=P+P_{GDE}$ and
\begin{eqnarray}\label{8}
\rho_{GDE}&=&\frac{1}{f_{\mathrm{R}}}\bigg[\big((1-
f_{\mathrm{R}})+\rho f_{\mathrm{T^2}}\big)\rho+
\frac{1}{2}(f-\mathrm{R}f_{\mathrm{R}})
-3H\dot{\mathrm{R}}f_{\mathrm{R}\mathrm{R}}\bigg],
\\\label{9}
P_{GDE}&=&\frac{1}{f_{\mathrm{R}}}\bigg[
\frac{1}{2}(\mathrm{R}f_{\mathrm{R}}-f)
+\ddot{\mathrm{R}}f_{\mathrm{R}\mathrm{R}}+
\dot{\mathrm{R}}^2f_{\mathrm{R}\mathrm{R}\mathrm{R}}+
2H\dot{R}f_{\mathrm{R}\mathrm{R}}\bigg].
\end{eqnarray}
Addition of Eqs.(\ref{8}) and (\ref{9}) gives the evolution equation
as
\begin{equation}\label{10}
\dot{R}^2 f_{\mathrm{R}\mathrm{R}\mathrm{R}}+(\ddot{\mathrm{R}}-
H\dot{\mathrm{R}})f_{\mathrm{R}\mathrm{R}}+(\rho f_{\mathrm{T^2}}+1
)\rho-((1+\omega_{DE})\rho_{DE}+\rho)f_{\mathrm{R}}=0.
\end{equation}
This is a third-order nonlinear partial differential equation in
$f(\mathrm{R})$ that encompasses both geometric and matter terms.
The non-conserved stress-energy tensor in this modified proposal
indicates the presence of geodesic motion of the particles which
yields additional physical effects that are not accounted in
standard models. This deviation shows that the energy exchange
between different sectors of the universe, gravitational
interactions with exotic matter or the influence of higher-order
curvature terms, provide a more comprehensive description of
gravitational dynamics \cite{7}. The corresponding non-conserved
stress-energy tensor turns out to be
\begin{equation}\label{11}
\nabla^{\mu} T_{\mu\nu}= \nabla^{\mu}\Theta_{\mu\nu}f_{\mathrm{T^2}}
-\frac{1}{2}g_{\mu\nu}\nabla^{\mu}f(\mathrm{T^2}).
\end{equation}
Using the standard continuity equation
$(\dot{\rho}+3H\rho(1+\omega)=0)$, we get
\begin{equation}\label{11a}
(3\omega^4+4\omega^2+1)f_{\mathrm{T^2}\mathrm{T^2}}+(6\omega^{4}+11\omega^{2}+3)f_{\mathrm{T^2}}=0.
\end{equation}

In the following section, we use two different functional forms in
the background of Garcia-Salcedo GDE and GGDE models to reformulate
the $f(\mathrm{R},\mathrm{T^2})$ gravity models.

\section{Reconstruction of $\mathrm{R}+2f(\mathrm{T^2})$ Gravity Model}

The field equations of this modified theory are in complex form
because of multi-variable functions and their derivatives. As a
result, we assume a specific $f(\mathrm{R},\mathrm{T^2})$ model as
\cite{8}
\begin{equation}\label{13}
f(\mathrm{R},\mathrm{T^2})= \mathrm{R}+2f(\mathrm{T^2}).
\end{equation}
The field equations (\ref{8}) and (\ref{9}) corresponding to this
model are
\begin{eqnarray}\label{15}
\rho_{GDE}&=&2\mathrm{T^2}f_{\mathrm{T^2}}(\mathrm{T^2})+f(\mathrm{T^2}),
\\\label{15a}
P_{GDE}&=&-f(\mathrm{T^2}).
\end{eqnarray}
The EoS parameter describes the relationship between pressure and
energy density in a given cosmological model, expressed as
$\omega=\frac{p}{\rho}$. In cosmological models involving modified
gravity theories like $f(\mathrm{R},\mathrm{T}^2)$, the EoS
parameter is crucial for understanding the dynamics of the universe.
The behavior of the EoS parameter can vary depending on the specific
form of the function and the matter content of the universe. In
cosmology, the ``phantom" epoch refers to a period when $\omega<-1$,
indicating that the energy density increases with expansion. On the
other hand, the ``quintessence" epoch refers to a phase when
$-1<\omega<-1/3$, characterizing a type of DE that leads to
accelerated expansion. The transition between these epochs, from
phantom to quintessence-like behavior depends on the specific
dynamics of the $f(\mathrm{R},\mathrm{T}^2)$ gravity model. It
involves a change in the effective behavior of DE as the universe
evolves. This transition could be driven by changes in the
functional form, the matter content of the universe, or both. The
EoS parameter of GDE $(\omega_{GDE}=\frac{P_{GDE}}{\rho_{GDE}})$
corresponding to $f(\mathrm{T^2})$ is computed as
\begin{equation}\label{16}
\omega_{GDE}=-\frac{f(\mathrm{T^2})}{2\mathrm{T^2}f_{\mathrm{T^2}}+f(\mathrm{T^2})}.
\end{equation}

Now, we rebuild $f(\mathrm{R},\mathrm{T^2})$ model through
Gracia-Salcedo GDE to investigate the dynamics of DE in the context
of modified gravity in the given subsection.

\subsection{Garcia-Salcedo GDE Model}

This model incorporates ghost fields inspired by QCD, offering a
novel perspective on the nature of DE. The behavior of DE and
gravitational interactions in this framework is determined by
trapping horizons, which are regions from which light cannot escape.
Using trapping horizons in the QCD model, researchers discussed the
relationship between DE dynamics and spacetime structures
\cite{30a}. This approach offers insights into the relationship
between quantum field theory and gravitational phenomena, presenting
a promising avenue for solving the cosmological puzzles. The GDE
model of QCD related to trapping horizon for non-zero constant
$\eta$ and curvature parameter $\mathcal{K}$ is given by \cite{31}
\begin{eqnarray}\label{17}
\rho_{GDE}=\eta(1-\varepsilon)\sqrt{H^{2}+\frac{\mathcal{K}}{{a}^2}},
\quad \varepsilon= \frac{\dot{\tilde{r}}}{2 H
\tilde{r}_{\mathrm{T}}}.
\end{eqnarray}
For the flat FRW model $(\mathcal{K}=0)$, the trapping and Hubble
horizons are considered as $\tilde{r}_{\mathrm{T}}=\frac{1}{H}$ and
$\varepsilon=-\frac{\dot{H}}{2H^2}$, respectively. Inserting these
values in Eq.(\ref{17}), we obtain
\begin{equation}\label{18}
\rho_{GDE}=\eta\bigg(1+\frac{\dot{H}}{2 H^{2}}\bigg)H.
\end{equation}
Using the standard continuity equation, the EoS parameter for DE
becomes
\begin{equation}\label{19}
1+\omega_{GDE}=\frac{1}{3}\bigg(\frac{\dot\varepsilon}{H(1-\varepsilon)}+
2\varepsilon\bigg).
\end{equation}
Using Eqs.(\ref{11a}) and (\ref{16}) in (\ref{19}), we obtain
\begin{equation}\label{20}
4(\mathrm{T^2})^2f_{\mathrm{T^2}\mathrm{T^2}}+3\bigg[1-\frac{1}{3}
\left(\frac{\dot{\varepsilon}}{1-\varepsilon H}+2
\varepsilon\right)^{-1}-1\bigg]f({\mathrm{T^2}})=0.
\end{equation}

This equation is not explicitly defined in terms of $\mathrm{T^2}$
which gives complexity in the derivation of an analytic solution for
the function $f(\mathrm{T^2})$ in QCD GDE model. However, an
alternative approach involves reconstructing the scale factor which
serves as a key component in determining this functional form
$f(\mathrm{T^2})$. For this purpose, we consider the scale factor as
\cite{32}
\begin{eqnarray}\label{22}
a(\mathrm{t})=a_{\circ}(\mathrm{t}_{s}-\mathrm{t})^{-m},\quad
m,a_{\circ}>0.
\end{eqnarray}
This model portrays the cosmic phantom regime that leads to the type
I singularity. Substituting the above equation and using the
relation for Hubble horizon into Eq.(\ref{20}), we obtain
\begin{equation}\label{23}
4m(\mathrm{T^2})^2f_{\mathrm{T^2}\mathrm{T^2}}+f({\mathrm{T^2}})=0.
\end{equation}
Solving this equation, we have
\begin{equation}\label{24}
f(\mathrm{T^2})=\alpha\mathrm{(T^2)}^{1-\sqrt{1-a}}+\beta\mathrm{(T^2)}^{1+\sqrt{1-a}},
\end{equation}
where $a={\frac{1}{m}}$ and $\alpha$, $\beta$ are the integration
constants. The right and left panels of Figure \textbf{1} show the
evolution of $f(\mathrm{T^2})$ against $\mathrm{T^2}$ and $1+z$,
respectively for various values of $m$, indicating that the
$f(\mathrm{T^2})$ is increasing for both the scenarios as required.

The variation of EoS parameter $\omega_{GDE}$ in the context of
$f(\mathrm{R},\mathrm{T^2})$ gravity is instrumental in delineating
the stages of cosmic expansion. For $\omega_{GDE}>-1$, marking the
quintessence era, the universe avoids transitions into either the de
Sitter or big rip phases. Quintessence signifies an epoch where
cosmic expansion accelerates, albeit at a gradually diminishing
rate, attributed to a dynamic form of DE. In contrast,
$\omega_{GDE}<-1$ characterizes the phantom epoch, wherein the null
energy condition is violated, potentially leading to a catastrophic
big rip scenario, with the universe expanding at an increasingly
rapid pace. At $\omega_{GDE}=-1$ representing the de Sitter phase,
the universe undergoes exponential expansion, mirroring a constant
DE density reminiscent of a cosmological constant. In the specific
model under scrutiny, within the framework of
$f(\mathrm{R},\mathrm{T^2})$ gravity theory, the observed
$\omega_{GDE}>-1$ indicates quintessence era of the universe. This
suggests that cosmic expansion is accelerating, yet without the
looming prospect of a big rip, aligning with a dynamic DE component
akin to quintessence as depicted in the Figure \textbf{2}. This
understanding aids in characterizing the ongoing phase of cosmic
evolution and its projected trajectories.
\begin{figure}
\epsfig{file=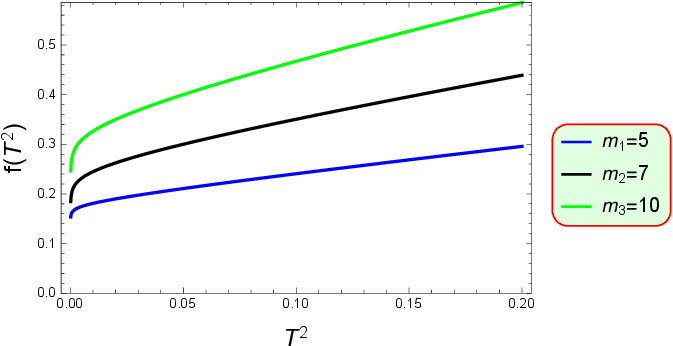,width=.5\linewidth}
\epsfig{file=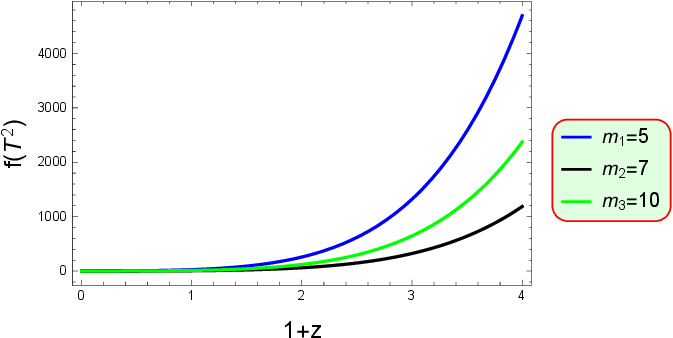,width=.5\linewidth}\caption{Evolution of
$f(\mathrm{T^2})$ versus $\mathrm{T^2}$ (left) and $1+z$ (right) for
considered values of the parameters.}
\end{figure}
\begin{figure}
\begin{center}
\epsfig{file=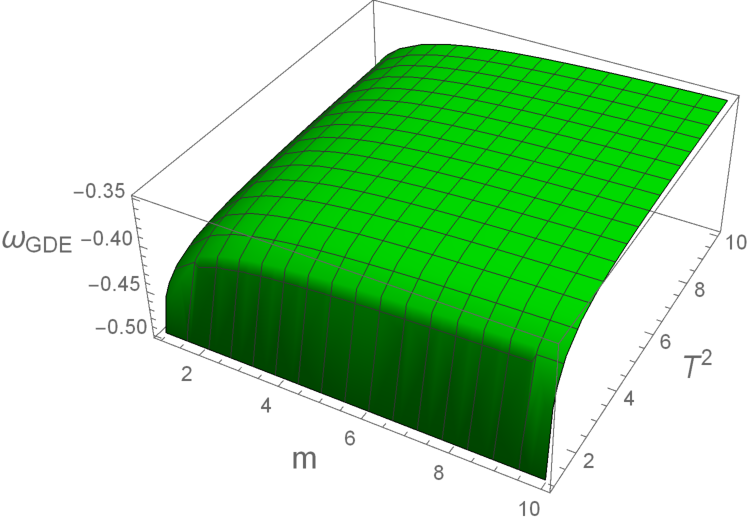,width=.5\linewidth} \caption{Graph of EoS
parameter versus $\mathrm{T}^2$ and $m$.}
\end{center}
\end{figure}

Now, we examine the stability of the reconstructed Garcia-Salcedo
GDE $f(\mathrm{R},\mathrm{T^2})$ model through linear homogeneous
perturbations. By analyzing how the model responds to such
perturbations, we can assess its viability as a theoretical
framework to determine the dynamics of the universe. Furthermore,
the stability allows us to identify the conditions under which the
model solutions remain consistent and reliable, providing valuable
insights to explain observational phenomena such as the accelerated
cosmic expansion. In this perspective, we assume
$H(\mathrm{t})=H_{h}$ \cite{26} and the corresponding energy density
becomes
\begin{equation}\label{28}
\rho_{h}=\rho_{\circ}e^{-3\int H_{h}d\mathrm{t}},
\end{equation}
where $\rho_{\circ}$ is the constant energy density. The perturbed
energy density and Hubble parameter are given by \cite{26}
\begin{eqnarray}\label{29}
\rho=\rho_{h}(1+\xi_{m}), \quad H=H_{h}(1+\xi).
\end{eqnarray}
We can expand the function $f(\mathrm{T^2})$ as
\begin{equation}\label{30}
f(\mathrm{T^2})=f^{h}+f^{h}_{\mathrm{T^2}}
(\mathrm{T^2}-(\mathrm{T^2}_{h})^{\frac{1}{2}})+\mathcal{O}^2,
\end{equation}
where $h$ indicates that the functions and their derivatives are
evaluated in accordance with the solution $H= H_h$ and
$\mathcal{O}^2$ term involves the higher power of $\mathrm{T^2}$.
Inserting Eqs.(\ref{29}) and (\ref{30}) into (\ref{7d}), we have
\begin{equation}\label{31}
\big(\mathrm{T^2}_{h}+4\mathrm{T^2}_{h}f^{h}_{\mathrm{T^2}}+
4(\mathrm{T^2}_{h})^{\frac{3}{2}}f^{h}_{\mathrm{T^2}\mathrm{T^2}}\big)\zeta_{m}=
6H^{2}_{h}\zeta.
\end{equation}
The second perturbation equation for the conserved energy-momentum
tensor is
\begin{equation}\label{32}
\dot{\zeta}_{m}+3H_{h}\zeta=0.
\end{equation}
Using the above two equations, we obtain
\begin{equation}\label{33}
\dot{\zeta}_{m}+\frac{1}{2H_{h}}
\big(\mathrm{T^2}_{h}+4\mathrm{T^2}_{h}f^{h}_{\mathrm{T^2}}+
4(\mathrm{T^2}_{h})^{\frac{3}{2}}f^{h}_{\mathrm{T^2}\mathrm{T^2}}\big)\zeta_{m}=0.
\end{equation}
Solving this equation, we have
\begin{equation}\label{33}
\zeta_{m}=\gamma e^{-\frac{1}{2}\int\tau\mathrm{T^2}d\mathrm{t}},
\end{equation}
where $\tau\mathrm{T^2}=\frac{\mathrm{T^2}_{h}}{H_{h}}
\big(1+4f^{h}_{\mathrm{T^2}}+4\mathrm{T^2}_{h}f^{h}_{{\mathrm{T^2}}\mathrm{T^2}}\big)$.
Combining Eqs.(\ref{31}) and (\ref{33}), we obtain
\begin{equation}\label{34}
\zeta=\frac{1}{6H_{h}}\gamma\tau(\mathrm{T^2})^2e^{-\frac{1}{2}\int\tau(\mathrm{T^2})^2d\mathrm{t}}.
\end{equation}
\begin{figure}
\begin{center}
\epsfig{file=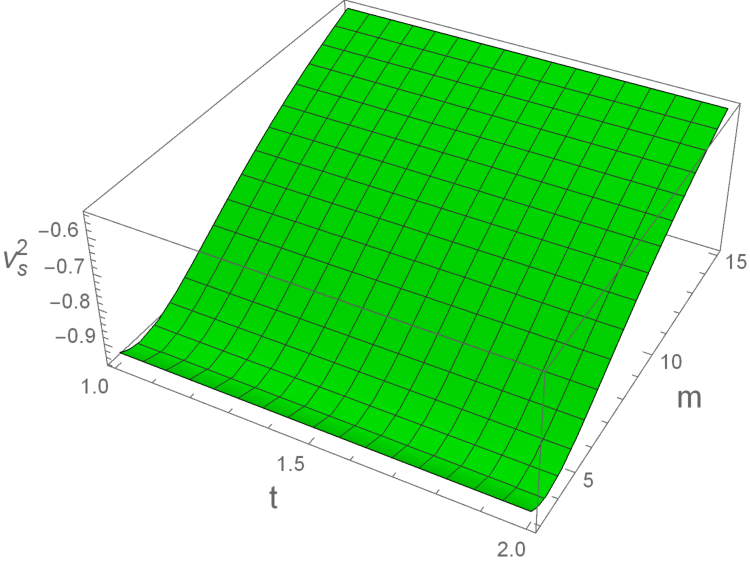,width=.5\linewidth} \caption{Plot of the
squared sound speed for $\mathrm{t}$ and $m$.}
\end{center}
\end{figure}

The stability of the reconstructed Garcia-Salcedo GDE
$f(\mathrm{R},\mathrm{T^2})$ gravity model (\ref{24}) is obtained by
analyzing the associated connections between $\tau T^2$ and
$-\frac{1}{2}\int\tau \mathrm{T^2}$ as
\begin{eqnarray}\nonumber
\tau\mathrm{T^2}&=&\mathrm{T^2}_{\circ}(\mathrm{t}_{s}-\mathrm{t})^{6m}+\mathrm{T^2}_{1}
\zeta(\mathrm{t}_{s}-\mathrm{t})^{6m(\zeta+1)-7}\\ \label{34a}
&+&\mathrm{T^2}_{2}\zeta(\zeta-1)
(\mathrm{t}_{s}-\mathrm{t})^{6m(\zeta+1)-13},
\\ \nonumber
-\frac{1}{2}\int\tau\mathrm{T^2}&=&\frac{1}{12m+2}\mathrm{T^2}_{\circ}(\mathrm{t}_{s}-t)^{6m+1}
+\zeta\mathrm{T^2}_{1}\frac{(\mathrm{t}_{s}-\mathrm{t})^{6(\zeta+m-1)}}{6(\zeta+m-1)}\\
\label{34b}&+&\mathrm{T^2}_{2}\zeta(\zeta-1)\frac{(\mathrm{t}_{s}-\mathrm{t})^{6(\zeta+m-2)}}{6(\zeta+m-2)},
\end{eqnarray}
where $\mathrm{T^2}_{\circ}=\rho^{2}_{\circ},~
\mathrm{T^2}_{1}=4\alpha\rho^{2\zeta}_{\circ}$ and
$\mathrm{T^2}_{2}=4\alpha\rho_{\circ}^{2\zeta+1}$. It is observed
that the above equations cannot diminish with time in future
evolution, consequently, leading to the instability of our model for
homogeneous perturbations. Thus the reconstructed Garcia-Salcedo GDE
$f(\mathrm{R},\mathrm{T}^2)$ gravity model displays instability
which coincides with findings in \cite{22a,26}. We  also examine the
evolution of the squared sound speed against $\mathrm{t}$ and $m$ in
Figure \textbf{3}, which shows that the corresponding model is
unstable as $\nu_{s}^2$ is negative.

\subsection{Generalized Ghost Dark Energy Model}
\begin{figure}
\begin{center}
\epsfig{file=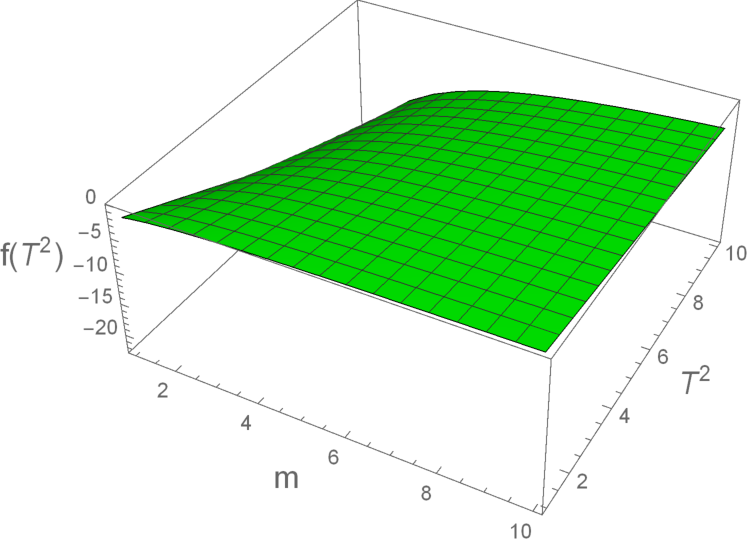,width=.5\linewidth} \caption{Plot of
$f(\mathrm{T^2})$ against $\mathrm{T^2}$ and $m$.}
\end{center}
\end{figure}

The GGDE model is defined through the Veneziano ghost field and the
expression for energy density of this model is given by \cite{3}
\begin{equation}\label{35}
\rho_{GDE}=\alpha H+\beta H^{2},
\end{equation}
where $\alpha$ and $\beta$ are nonzero constants. The EoS parameter
for this model is
\begin{equation}\label{35}
1+\omega_{GDE}=-\frac{1}{3}\bigg(\frac{\alpha+2\beta H}{\alpha+
\beta H}\bigg).
\end{equation}
Combining the above equation with Eq.(\ref{16}), we get
\begin{equation}\label{36}
f(\mathrm{T^2})=\frac{6m\big(\alpha+\frac{\beta\big(-\alpha
-\sqrt{\alpha^2-4\beta\mathrm{T^2}+12\mathrm{T^2}}\big)}{2(\beta
-3)}\big)}{\alpha+\mathrm{T^2}\big(\alpha+\frac{\beta\big(-\alpha
-\sqrt{\alpha^2-4\beta\mathrm{T^2}+12\mathrm{T^2}}\big)}{\beta
-3}\big)+\frac{\beta\big(-\alpha-\sqrt{\alpha^2-4\beta\mathrm{T^2}+12
\mathrm{T^2}}\big)}{\beta -3}}.
\end{equation}
Figure \textbf{4} illustrates the function $f(\mathrm{T^2})$ plotted
against $\mathrm{T^2}$ and $m$. The negative behavior of
$\mathrm{T^2}$ shows that the functional form is not viable and may
lead to matter-dominated era. This is confirmed through the EoS
parameter. Inserting Eq.(\ref{36}) into (\ref{16}), we obtain the
EoS parameter as
\begin{eqnarray}\nonumber
\omega_{GDE}&=&-\bigg[(\mathrm{T^2}+1)\sqrt{\alpha^2-4(\beta-3)
\mathrm{T^2}}\big(3\alpha+\beta\sqrt{\alpha^2-4(\beta-3)\mathrm{T^2}}\big)
\\ \nonumber
&\times&\big(\beta \sqrt{\alpha^2-4(\beta-3)\mathrm{
T}}-\alpha(\beta-6)\big)\bigg]\bigg[4(\beta-3)\beta
({\mathrm{T^2}})^2\big(\beta
\\ \nonumber&\times&
\sqrt{\alpha^2-4(\beta-3)\mathrm{T^2}}-2\alpha(\beta
-6)\big)+\alpha^2\big(((\beta-3)\beta+18)\\
\nonumber&\times&\sqrt{\alpha^2-4(\beta
-3)\mathrm{T^2}}-\alpha(\beta-9)\beta\big)+\mathrm{T^2}
\big(\alpha^3(\beta-9)\beta-24\\ \nonumber&\times&\alpha(\beta
-3)\beta-4(\beta-3)\beta ^2 \sqrt{\alpha
^2-4(\beta-3)\mathrm{T^2}}-\alpha^2((\beta-3)\beta
\\\label{37}&+&18)
\sqrt{\alpha^2-4(\beta-3)\mathrm{T^2}}\big)\bigg]^{-1}.
\end{eqnarray}
The variation of the EoS parameter in Figure \textbf{5} indicates
that the reconstructed model represents matter-dominated era as
$\omega_{GDE}>0$, hence is not viable.
\begin{figure}
\begin{center}
\epsfig{file=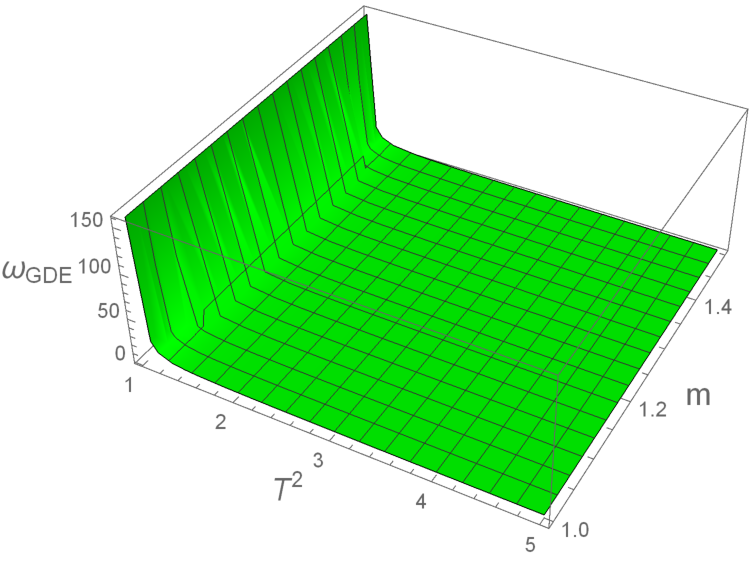,width=.5\linewidth} \caption{Evolution of EoS
versus $\mathrm{T^2}$ and $m$.}
\end{center}
\end{figure}

\section{Reconstruction of $f(\mathrm{R},\mathrm{T^2})=f_{1}(\mathrm{R})+f_{2}(\mathrm{T^2})$ Gravity Model}

This model stems from the desire to explore novel theoretical
frameworks that can effectively describe the cosmic dynamics. This
enables us to study the interplay between geometry and matter
content in a unified manner, potentially providing deeper insights
into cosmological phenomena such as cosmic acceleration and
gravitational dynamics on both micro and macro scales. Moreover,
this approach offers a significant platform to examine various
modifications to GR, which may better align with observational data
and address the cosmological puzzles. Inserting this functional form
into Eq.(\ref{10}), we obtain the evolution equation as
\begin{equation}\label{38}
\dot{\mathrm{R}}^2
f_{1\mathrm{R}\mathrm{R}\mathrm{R}}+(\ddot{\mathrm{R}}-
H\dot{\mathrm{R}})f_{1\mathrm{R}\mathrm{R}}+(\rho
f_{2\mathrm{T^2}}+1
)\rho-((1+\omega_{DE})\rho_{DE}+\rho)f_{1\mathrm{R}}=0.
\end{equation}
To maintain consistency between
$f(\mathrm{R},\mathrm{T^2})=f_{1}(\mathrm{R})+f_{2}(\mathrm{T^2})$
and the standard continuity equation, the right-hand side of
Eq.(\ref{10}) must be zero. Thus, we obtain
\begin{equation}\label{39}
f_{2\mathrm{T^2}}(6\omega^{4}+11\omega^{2}+3)+2\mathrm{T^2}
f_{2\mathrm{T^2}\mathrm{T^2}}(3\omega^{4}+4\omega^{2}+1)=0.
\end{equation}
The solution of this equation yields
\begin{equation}\label{40}
f_{2\mathrm{T^2}}=\beta_{1}\mathrm{T^2}^{\frac{3\omega^{2}+1}
{-2(3\omega^{4}+4\omega^{2}+1)}}+\beta_{2}.
\end{equation}
Here, $\beta_1$ and $\beta_2$ are constants and this specific
functional form satisfies the standard continuity equation. To
reformulate $f(\mathrm{R},\mathrm{T^2})$ model according to
Garcia-Salcedo and GGDE, we assume
$f_{2\mathrm{T^2}}=\beta_3\mathrm{T^2}$ ($\beta_3$ is an arbitrary
constant) as \cite{23}
\begin{equation}\label{40}
f(\mathrm{R},\mathrm{T^2})=f_{1}(\mathrm{R})+\beta_3\mathrm{T^2}.
\end{equation}
Putting this model in Eq.(\ref{38}), we have
\begin{equation}\label{41}
\dot{\mathrm{R}}^2
f_{1\mathrm{R}\mathrm{R}\mathrm{R}}+(\ddot{\mathrm{R}}-
H\dot{\mathrm{R}})f_{1\mathrm{R}\mathrm{R}}+(\rho\beta_3+1
)\rho-((1+\omega_{DE})\rho_{DE}+\rho)f_{1\mathrm{R}}=0,
\end{equation}
where
\begin{eqnarray}\label{42}
\dot{\mathrm{R}}^2=\frac{2\mathrm{R}^3}{6m(1+2m)}, \quad
\ddot{\mathrm{R}}-H\dot{\mathrm{R}}=
\frac{(3-2m)}{6m(1+2m)}\mathrm{R}^2.
\end{eqnarray}

\subsection{Garcia-Salcedo GDE Model}

First, we explore the Garcia-Salcedo GDE model. Using the GDE model
(\ref{17}), we obtain
\begin{eqnarray}\label{43}
&&\rho=\frac{m
\mathrm{R}}{(2+4m)}(1-\frac{(1+2\eta\sqrt{m})\sqrt{1+2m}}{m\sqrt{6m\mathrm{
R}}}),\\\label{44}
&&\rho_{GDE}(1+\omega_{GDE})=-\frac{(2\eta\sqrt{m}+1)\mathrm{R}^{\frac{3}{2}}}{6m\sqrt{6m+12m^2}}.
\end{eqnarray}
Putting Eqs.(\ref{42})-(\ref{44}) into (\ref{41}) yields
\begin{eqnarray}\nonumber
&&\mathrm{R}^3f_{1\mathrm{R}\mathrm{R}\mathrm{R}}+\big(\frac{3-2m}{2}\big)
\mathrm{R}^2f_{1\mathrm{R}\mathrm{R}}-m(1-f_{\mathrm{R}})+\frac{9\beta_3
m^{4}}{4}\mathrm{R}^2
\\\nonumber
&&+\frac{\beta_3
(1+2\eta\sqrt{m})^2(3m(1+2m))}{2}\mathrm{R}+\bigg[\sqrt{\frac{1+2m}{m}}(2\eta\sqrt{m}\\
\label{45}&&+1)-\frac{3\beta_3 m^2
(1+2\eta\sqrt{m})(\sqrt{3m(1+2m)})}{\sqrt{2}}\bigg]\mathrm{R}^{\frac{3}{2}}=0.
\end{eqnarray}
Its solution is given as
\begin{eqnarray}\nonumber
f_{1}(\mathrm{R})&=&\frac{a_{1}\big(2m+\sqrt{4(m-5)m+1}+3\big)
\mathrm{R}^{\frac{1}{4}\big(2m-\sqrt{4(m-5)m+1}+3\big)}}{8 m+2}\\
\nonumber&-&\frac{a_{2}\big(-2
m+\sqrt{4(m-5)m+1}-3\big)\mathrm{R}^{\frac{1}{4}\big(2m+\sqrt{4
(m-5)m+1}+3\big)}}{8 m+2}\\\nonumber&-&\frac{3}{4} m^4 \mathrm{R}^2
\beta_3 -\frac{2\mathrm{R}^{3/2}\big(2\eta\sqrt{m}+1\big)\big(2
\sqrt{\frac{1}{m}+2}-3\sqrt{6}\sqrt{m(2 m+1)}\beta_3\big)} {3
(m+1)}\\\label{46}&-&\frac{3}{2}(2
m+1)\mathrm{R}\beta_3\big(2\eta\sqrt{m}+1\big)^2+\mathrm{R}+a_{3},
\end{eqnarray}
where $a_{1}$, $a_{2}$ and $a_{3}$ are constants. The reconstructed
Garcia-Salcedo GDE model is obtained by putting the above equation
in Eq.(\ref{40}) as
\begin{eqnarray}\nonumber
f(\mathrm{R},\mathrm{T^2})&=&\frac{a_{1}\big(2m+\sqrt{4(m-5)m+1}+3\big)
\mathrm{R}^{\frac{1}{4}\big(2m-\sqrt{4(m-5)m+1}+3\big)}}{8 m+2}\\
\nonumber&-&\frac{a_{2}\big(-2
m+\sqrt{4(m-5)m+1}-3\big)\mathrm{R}^{\frac{1}{4}\big(2m+\sqrt{4
(m-5)m+1}+3\big)}}{8 m+2}\\\nonumber&-&\frac{3}{4} m^4\mathrm{R}^2
\beta_3-\frac{2\mathrm{R}^{3/2}\big(2\eta\sqrt{m}+1\big)\big(2
\sqrt{\frac{1}{m}+2}-3\sqrt{6}\sqrt{m(2 m+1)}\beta_3\big)}{3
(m+1)}\\\label{47}&-&\frac{3}{2}(2
m+1)\mathrm{R}\beta_3\big(2\eta\sqrt{m}+1\big)^2+\mathrm{R}+\beta_3\mathrm{T^2}+a_{3}.
\end{eqnarray}
The graphical evolution of the reconstructed Garcia-Salcedo GDE
$f(\mathrm{R},\mathrm{T^2})$ model versus redshift parameter
$z=a^{-1}(1-a)$ is depicted in Figure \textbf{6}. We have selected
the values of free parameters as $a_{1}=-1,~a_{2}=-0.25$,
$a_{3}=-10,~ \eta=1.5,~\beta_3=0.5$ with $m$ = 5, 7, 10. The
reconstructed model exhibits a gradual increase with the increase in
redshift parameter representing the accelerated expansion of the
cosmos. Additionally,
\begin{equation}\label{46.1}
\lim_{z\rightarrow0}f(\mathrm{R},\mathrm{T^2})=0,
\end{equation}
which suggests that the reconstructed Garcia-Salcedo GDE model is
realistic. The corresponding expressions for energy density and
pressure turn out to
\begin{eqnarray}\label{46.2}
\rho_{GDE}&=&\frac{1}{f_{\mathrm{R}}}\bigg[((1-
f_{\mathrm{R}})\mathrm{T^2}+\beta_3\mathrm{T^2})+
\frac{1}{2}(f-\mathrm{R}f_{\mathrm{R}})
-3H\dot{\mathrm{R}}f_{\mathrm{R}\mathrm{R}}\bigg],
\\\label{46.3}
P_{GDE}&=&\frac{1}{f_{\mathrm{R}}}\bigg[
\frac{1}{2}(\mathrm{R}f_{\mathrm{R}}-f)-\beta_3 \mathrm{T^2})
+\ddot{\mathrm{R}}f_{\mathrm{R}\mathrm{R}}+
\dot{\mathrm{R}}^2f_{\mathrm{R}\mathrm{R}\mathrm{R}}+
2H\dot{R}f_{\mathrm{R}\mathrm{R}}\bigg].
\end{eqnarray}
\begin{figure}
\begin{center}
\epsfig{file=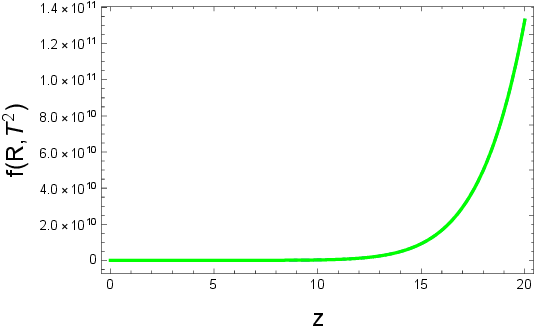,width=.5\linewidth} \caption{Evolution of
Garcia-Salcedo GDE $f(\mathrm{R},\mathrm{T^2})$ model for different
values of $m$.}
\end{center}
\end{figure}

The null energy condition (NEC) plays a crucial role in cosmology,
where it serves as a fundamental criterion for the viability of
certain DE models. In this context, NEC imposes restrictions on the
energy density and pressure of the fluid. Violation of the NEC
indicates the exotic matter which leads to unconventional
cosmological scenarios such as phantom energy. Understanding and
exploring the implications of the violation of NEC is essential for
unraveling the mysteries of DE and its role in cosmic acceleration.
We study the graphical evolution of the NEC and EoS parameter for
Garcia-Salcedo GDE $f(\mathrm{R},\mathrm{T^2})$ model. The panel on
the left hand side of Figure \textbf{7} shows that the NEC is
violated ($\rho_{GDE}+P_{GDE}<0)$, whereas the right panel reveals
that the EoS is less than -1, i.e., $(\omega_{GDE}<-1)$. Both the
conditions favor the cosmic phantom regime. The behavior of squared
speed of sound $(\nu_{s}^2=\frac{\dot{P}_{GDE}}{\dot{\rho}_{GDE}})$
is demonstrated in Figure \textbf{8}, which shows that the
reconstructed Garcia-Salcedo GDE $f(\mathrm{R},\mathrm{T^2})$ model
is unstable.
\begin{figure}
\epsfig{file=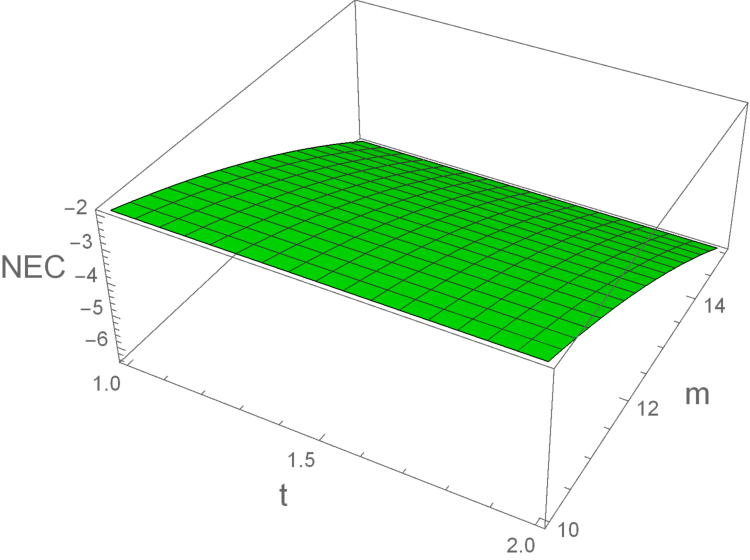,width=.5\linewidth}
\epsfig{file=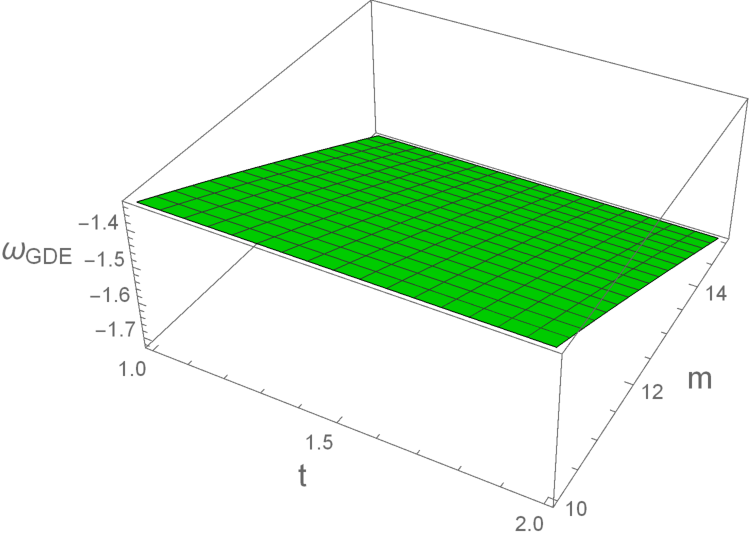,width=.5\linewidth}\caption{Behavior of NEC (left)
and EoS parameter (right) against $m$ and temperature.}
\end{figure}
\begin{figure}
\begin{center}
\epsfig{file=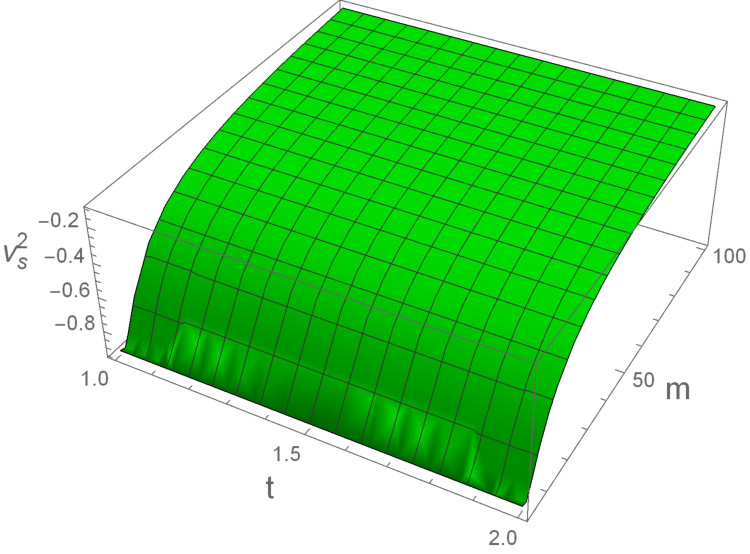,width=.5\linewidth} \caption{Variation of
squared sound speed against $m$ and  $\mathrm{t}$.}
\end{center}
\end{figure}

\subsection{Generalized Ghost Dark Energy Model}

In this section, we use GGDE model to reconstruct viable
$f(\mathrm{R},\mathrm{T}^{2})$ model. Using the GGDE model in
Eq.(\ref{35}), we have
\begin{eqnarray}\label{47}
&&\rho=\frac{m\mathrm{R}}{6+12m}\big((3-\beta)-\alpha\sqrt{\frac{6+12m}{m\mathrm{R}}}\big),
\\\label{48}
&&\rho_{GDE}(1+\omega_{GDE})=\frac{1}{\sqrt{m}(1+2m)}
\big(\alpha\sqrt{(6+12m)\mathrm{R}}+6\sqrt{m}\beta\mathrm{R}\big).
\end{eqnarray}
Inserting Eqs.(\ref{42}), (\ref{47}) and (\ref{48}) into (\ref{41}),
we obtain
\begin{eqnarray}\nonumber
&&\mathrm{R}^{3}f_{\mathrm{R}\mathrm{R}\mathrm{R}}+(\frac{3-2m}{2})
\mathrm{R}^2f_{\mathrm{R}\mathrm{R}}-m(1-f_{\mathrm{R}})\mathrm{R}
-6\alpha\sqrt{6m(1+2m)}\mathrm{R}^{\frac{1}{2}}
\\\label{49}&&-36m\beta\mathrm{R}+
\beta_3\frac{m^3}{6+12m}\big((3-\beta)-\alpha\sqrt{\frac{6(1+2m)}{m\mathrm{R}}}\big)^{2}\mathrm{R}^2=0.
\end{eqnarray}
The corresponding solution is
\begin{eqnarray}\nonumber
f(\mathrm{R})&=&\frac{c_1\big(2m+\sqrt{4(m-5)m+1}+3\big)
\mathrm{R}^{\frac{1}{4}\big(2 m-\sqrt{4(m-5)m+1}+3\big)}}{8 m+2}\\
\nonumber&-&\frac{c_2\big(-2 m+\sqrt{4 (m-5) m+1}-3\big)
\mathrm{R}^{\frac{1}{4}\big(2 m+\sqrt{4(m-5)m+1}+3\big)}}{8 m+2}\\
\nonumber&-&\frac{(\beta-3)\beta_3 m^3R^2\big(\beta+(\beta-3)
m+8\alpha\sqrt{\frac{12 m+6}{m R}}-3\big)}{18(m+1)(2 m+1)}\\
\label{50}&+&
\mathrm{R}\big(36\beta+\alpha^2\beta_3(-m)+1\big)+\frac{8\sqrt{6}\alpha\sqrt{m
(2 m+1)}\sqrt{\mathrm{R}}}{m}+c_3.
\end{eqnarray}
The reconstructed GGDE model is obtained as
\begin{eqnarray}\nonumber
f(\mathrm{R},\mathrm{T^2})&=&\frac{c_1\big(2m+\sqrt{4(m-5)m+1}+3\big)
\mathrm{R}^{\frac{1}{4}\big(2 m-\sqrt{4(m-5)m+1}+3\big)}}{8 m+2}\\
\nonumber&-&\frac{c_2\big(-2 m+\sqrt{4 (m-5) m+1}-3\big)
\mathrm{R}^{\frac{1}{4}\big(2 m+\sqrt{4(m-5)m+1}+3\big)}}{8 m+2}\\
\nonumber&-&\frac{(\beta-3)\beta_3 m^3R^2\big(\beta+(\beta-3)
m+8\alpha\sqrt{\frac{12 m+6}{m R}}-3\big)}{18(m+1)(2 m+1)}
+\mathrm{R}\big(36\beta\\
\label{50a}&+&\alpha^2\beta_3(-m)+1\big)+\frac{8\sqrt{6}\alpha\sqrt{m
(2 m+1)}\sqrt{\mathrm{R}}}{m}+\beta_3\mathrm{T^2}+c_3.
\end{eqnarray}
\begin{figure}
\begin{center}
\epsfig{file=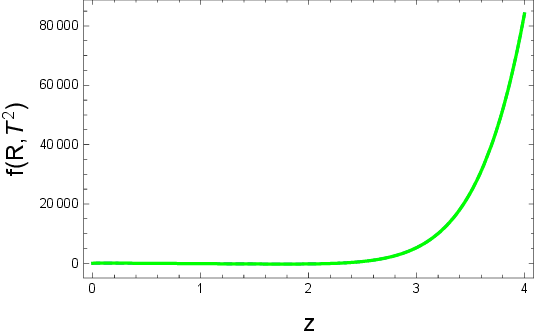,width=.5\linewidth} \caption{Plot of GGDE
$f(\mathrm{R},\mathrm{T^2})$ for considered values of $m$.}
\end{center}
\end{figure}

Figure \textbf{9} illustrates how the reconstructed GGDE
$f(\mathrm{R},\mathrm{T^2})$ model changes with respect to the
redshift parameter for $c_1=1$, $c_2=0.25$, $c_3=-10$, $\alpha=10$,
$\beta=0.5$, $\beta_3=0.5$ and $m=5, 7, 10$. It is found that the
reconstructed model increases gradually with the redshift parameter.
Further,
\begin{eqnarray}
\lim_{z\rightarrow0}f(\mathrm{R},\mathrm{T^2})=0,
\end{eqnarray}
which indicates that the reconstructed model is viable. We examine
the behavior of the NEC and EoS parameter for the reconstructed GGDE
model against $m$ and $t$. Figure \textbf{10} indicates that the NEC
is violated (left panel) and supports the phantom cosmic era as
verified from the EoS parameter (right panel). The stability of the
reconstructed GGDE model is checked through the squared speed of
sound. Figure \textbf{11} shows that the universe dominated by GGDE
is not stable.
\begin{figure}
\epsfig{file=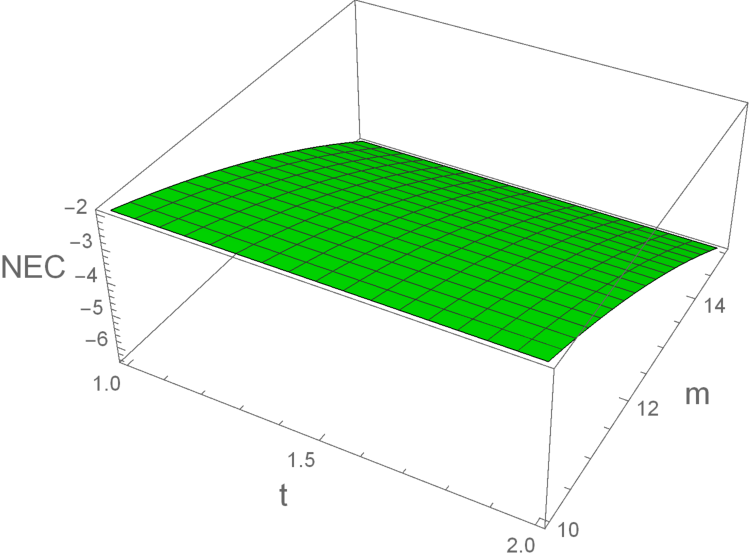,width=.5\linewidth}
\epsfig{file=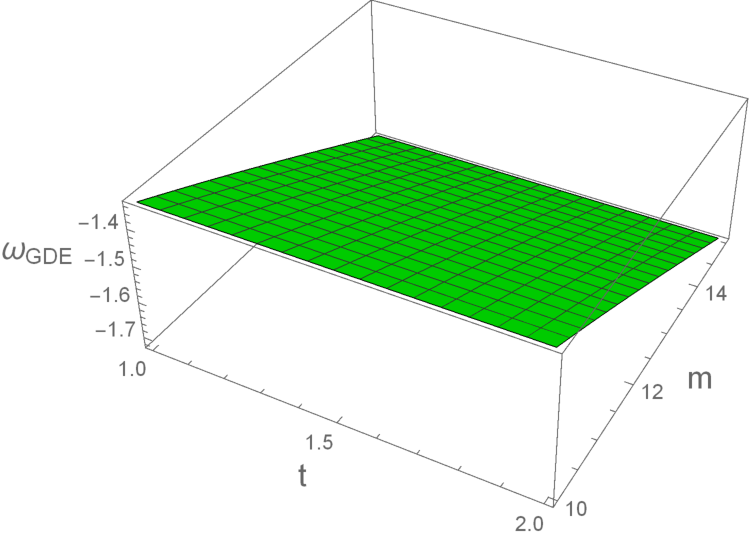,width=.5\linewidth}\caption{Null energy condition
(left) and EoS parameter (right) against $m$ and $\mathrm{t}$ for
considered values of the model parameters.}
\end{figure}
\begin{figure}
\begin{center}
\epsfig{file=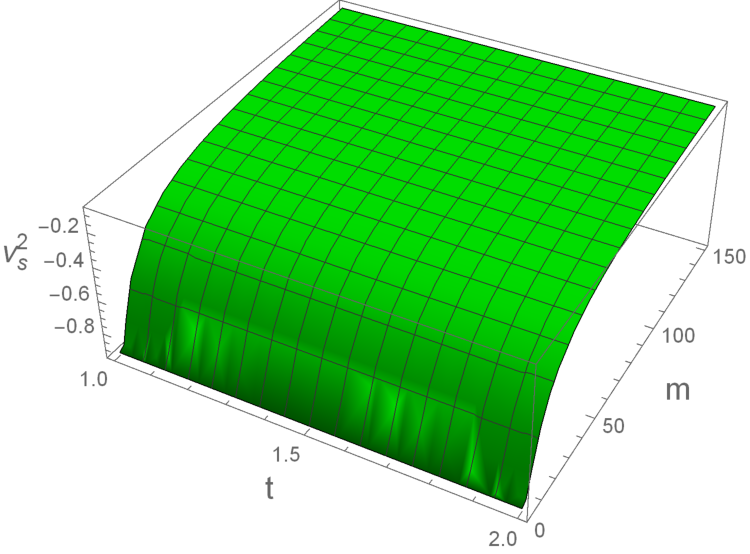,width=.5\linewidth} \caption{Squared sound
speed with respect to $m$ and  $\mathrm{t}$.}
\end{center}
\end{figure}

\section{Conclusions}

Recent scientific development involves modifying matter by adding a
term proportional to $\mathrm{T}^{\mu\nu}\mathrm{T}_{\mu\nu}$. This
deviation from linearity has significant consequences for the
dynamics of cosmology, especially in the high energy regime. It
indicates that interaction with matter itself becomes crucial,
particularly in the early stages of the universe. In vacuum, this
theory reduces to GR, with its effects primarily affecting the
distribution of matter and energy. Researchers are deeply intrigued
by the cosmological implications of this modified theory, which goes
beyond its relevance in the early universe and rebounding solutions.
This has shown its alignment with the solar system test, indicating
a broader scope of investigation. This paper studies the
reconstruction of the $f(\mathrm{R},\mathrm{T^2})$ gravity model
using the Gracia-Salcedo GDE and GGDE models to discuss the cosmic
evolution. The main focus is on the scale factor that represents the
phantom era of the universe and leads to a type I singularity
\cite{33}. The non-zero conservation of the stress-energy tensor is
a key concern in modified theories. To address this issue, the
explicit form of the functions $f(\mathrm{R})$ and $f(\mathrm{T^2})$
are derived by imposing the constraint of the standard continuity
equation.

The Garcia-Salcedo GDE, GGDE models and modified
$f(\mathrm{R},\mathrm{T}^2)$ theory each provide unique perspectives
and mechanisms that contribute to our understanding of cosmic
evolution and expansion. The DE models offer a mechanism for the
accelerated expansion of the universe without relying on a
cosmological constant. These models explain different phases of
cosmic evolution by adjusting the dynamics of ghost fields, leading
to transitions between decelerating and accelerating phases of the
universe. The GGDE model extends the basic idea of GDE by
incorporating additional parameters and functional forms to describe
the ghost field's potential and interactions. By generalizing the
ghost field dynamics, these models can produce a richer variety of
cosmic evolution scenarios, including different rates of
acceleration and transitions between different phases of cosmic
expansion. Modified $f(\mathrm{R},\mathrm{T}^2)$ theory encompasses
a broad range of models that explain cosmic acceleration. Thus,
Garcia-Salcedo GDE, GGDE models and modified gravitational theory
provide innovative frameworks that extend beyond the standard
cosmological model. They offer new ways to understand the
accelerated expansion of the universe and its cosmic evolution, each
bringing unique mechanisms and theoretical insights that help to
address some of the deepest questions in cosmology.

Here, we provide a summary of our findings for the two models.
\begin{itemize}
\item $f(\mathrm{R},\mathrm{T^2})=\mathrm{R}+2f(\mathrm{T^2})$
\end{itemize}
The $f(\mathrm{T^2})$ corresponding to the Garcia-Salcedo GDE
(\ref{24}) (Figure \textbf{1}) demonstrates increasing behavior
which favors the accelerated cosmic expansion and hence is in line
with observational data. The EoS parameter is greater than -1,
representing quintessence cosmic era (Figure \textbf{2}) and is
consistent with WMAP9 observations \cite{36}. We have also checked
the stability against linear homogeneous perturbations, which
reveals instability (Figure \textbf{3}). The reconstructed
$f(\mathrm{T^2})$ GGDE model indicates the decreasing behavior,
strongly suggesting that this functional form is not viable. The EoS
parameter (\ref{36}) represents the matter dominated era (Figure
\textbf{5}), hence this model lacks viability in the light of recent
observational data.
\begin{itemize}
\item $f(\mathrm{R},\mathrm{T^2})=f_{1}(\mathrm{R})+f_{2}(\mathrm{T^2})$
\end{itemize}
Here, the $f(\mathrm{R},\mathrm{T^2})$ gravity Garcia-Salcedo GDE
model shows monotonically increasing behavior as required (Figure
\textbf{6}). The graphical behavior of NEC and the EoS parameter
shows the cosmic phantom regime (Figure \textbf{7}). Hence, the
Garcia-Salcedo GDE $f(\mathrm{R},\mathrm{T^2})$ gravity model
supports the cosmic expansion. For the GGDE, the reconstructed model
indicates the fluctuations for both the NEC and the EoS parameter
(Figure \textbf{11}). Our findings align with the recent
observational data \cite{36}.\\\\
\textbf{Data Availability:} No data was used for the research
described in this paper.

\end{document}